\documentclass{webofc}

\usepackage[varg]{txfonts}   

\usepackage{hyperref}       
\usepackage{url}            
\usepackage{booktabs}       
\usepackage{amsfonts}       
\usepackage{nicefrac}       
\usepackage{microtype}      
\usepackage[dvipsnames]{xcolor}         

\usepackage{comments}

\hypersetup{
    colorlinks=true,
}

\usepackage{graphicx}
\usepackage{amsmath,amssymb,mathtools,amsthm}
\usepackage[capitalize,noabbrev]{cleveref}

\usepackage[bibliography=separate]{apxproof}

\usepackage{multiaudience}
\SetNewAudience{chep}
\SetNewAudience{thesis}
\DefCurrentAudience{chep}

\usepackage{mleftright}
\usepackage{subfigure}
\usepackage{tikz}

\usetikzlibrary{fit,positioning,shapes.geometric,decorations.text,decorations.pathreplacing,arrows,patterns}

\tikzstyle{alg} = [rectangle, minimum width=1.3cm, minimum height=0.8cm, text centered, text width=1cm, draw=black]
\tikzstyle{io} = [rectangle, minimum width=3cm, minimum height=1cm, text centered, text width=3cm, draw=black]
\tikzstyle{nn} = [rectangle, minimum width=3cm, minimum height=0.5cm, text centered, text width=6cm, draw=black]
\tikzstyle{arrow} = [thick,->,>=stealth]

\graphicspath{{Figures/}}

\newcommand{\Bernoulli}[1]{\text{Ber}\left(#1\right)}
\newcommand{\kl}[2]{D_{KL} \left(#1 || #2 \right)}
\newcommand{\sigmoid}[1]{\sigma \left(#1\right)}
\newcommand{\logit}[1]{\text{logit}\left(#1\right)}
\newcommand{\norm}[1]{\left\lVert#1\right\rVert}

\makeatletter
\newcommand*{\e}{%
  \def\e@sub{}%
  \e@scripts
}
\newcommand*{\e@scripts}{%
  \@ifnextchar_\e@subscript{%
    \e@finish
  }%
}
\def\e@subscript_#1{%
  \ifx\e@sub\@empty
    \def\e@sub{#1}%
  \else
    \errmessage{e already has a subscript}%
  \fi
  \e@scripts
}
\newcommand*{\e@finish}[1]{%
  \mathbb{E}%
  \ifx\e@sub\@empty\else _{\e@sub}\fi
  \mleft[#1\mright]%
}
\makeatother

\begin{document}
\title{Resilient VAE: Unsupervised Anomaly Detection at the SLAC Linac Coherent Light Source}
%
%

\author{
    \firstname{Ryan} \lastname{Humble}\inst{1} \fnsep \thanks{\email{ryhumble@stanford.edu}}
    \and
    \firstname{William} \lastname{Colocho}\inst{2}
    \and
    \firstname{Finn} \lastname{O'Shea}\inst{2}
    \and
    \firstname{Daniel} \lastname{Ratner}\inst{2}
    \and
    \firstname{Eric} \lastname{Darve} \inst{1,3}
}

\institute{
    Institute for Computational and Mathematical Engineering, Stanford University
    \and
    SLAC National Laboratory
    \and
    Department of Mechanical Engineering, Stanford University
}

\abstract{%
Significant advances in utilizing deep learning for anomaly detection have been made in recent years. However, these methods largely assume the existence of a normal training set (i.e., uncontaminated by anomalies) or even a completely labeled training set. In many complex engineering systems, such as particle accelerators, labels are sparse and expensive; in order to perform anomaly detection in these cases, we must drop these assumptions and utilize a completely unsupervised method.
\showto{thesis}{Moreover, only identifying the anomaly is insufficient: operators of these complex systems need additional localization information to identify the root cause of the anomaly and make an informed response.}
This paper introduces the Resilient Variational Autoencoder (ResVAE), a deep generative model specifically designed for anomaly detection. ResVAE exhibits resilience to anomalies present in the training data and provides feature-level anomaly attribution. During the training process, ResVAE learns the anomaly probability for each sample as well as each individual feature, utilizing these probabilities to effectively disregard anomalous examples in the training data. We apply our proposed method to detect anomalies in the accelerator status at the SLAC Linac Coherent Light Source (LCLS). By utilizing shot-to-shot data from the beam position monitoring system, we demonstrate the exceptional capability of ResVAE in identifying various types of anomalies that are visible in the accelerator.
}
\maketitle
\section{Introduction}
Anomaly detection (AD), which is the task of finding abnormal data or events, is a critical task for nearly all data-heavy, complex systems, such as industrial facilities, manufacturing, and large-scale science experiments
\showto{chep}{~\cite{Edelen21,Radaideh22,Fol2020,Valentino2017}}
\showto{thesis}{~\cite{Sun16,Zhao19,Lutz20,Edelen21,Lindemann21,Radaideh22,Fol2020,Dewitte19,Valentino2017}}.
These systems can produce thousands of real-time signals, which quickly overwhelm human operators that seek to monitor system performance. Identifying failures in these systems is a critical task, as failures can result in faulty outputs or cause damage to components. Unfortunately, the same complexity that overwhelms operators ensures that labeled data is rare or nonexistent and expensive to acquire.

In this work, we focus on detecting anomalies at SLAC's Linac Coherent Light Source (LCLS), which is a free-electron laser (FEL) system that enables users to take X-ray snapshots of microscopic phenomena. Since LCLS loses approximately 3\% of availability annually to unplanned downtime and experiences additional beam degradation without downtime, the task of identifying failures is critical to maximizing the amount of stable X-ray beam delivered to user experiments and thereby maximizing the scientific contribution of LCLS. Unfortunately, detecting anomalies at LCLS is challenging because of the data volume and lack of labeled data. 
\showto{thesis}{In particular, there are over 200,000 different process variable (PV) streams at LCLS (many of which record data at the full beam rate of $120$Hz), and none are systematically linked to the failures that operators do observe.}
This combination makes most traditional AD methods~\cite{Reynolds09_GMM, Parzen62_KDE, Scholkopf01_OCSVM,  Liu08} challenging to apply.
Deep learning approaches~\cite{Chen17,Lu17_OD_AE,Wang17_VAE,Su19_OmniAnomaly,Liu19_GAAL,Zong18_DAGMM,Ruff18_DSVDD}
are often quite effective in this high-volume, high-dimensional setting; however, they nearly always assume a normal-only training set. Since we lack labels, our training data will inevitably be contaminated by anomalies.

Our work's main contributions are: (i) introduction of a new deep generative model for anomaly detection, called Resilient VAE, to cope with training data contamination (\cref{sec:resvae}); and (ii) its application to identifying anomalies in accelerator status at LCLS (\cref{sec:lcls_application}).



\section{Resilient VAE}\label{sec:resvae}
We consider a particular type of anomaly detection task where the data is high-dimensional and completely unlabeled. We suppose a dataset $ \mathcal{D} = \{x\} $, where each sample is $ x \in \mathbb{R}^D $ and $D$ is the number of features. We suppose an (unknown) anomaly process that contaminates samples and features there-within, in our dataset. In particular, we assume that any training dataset has already been contaminated by anomalies, so there is no normal-only training set.

\begin{toappendix}
    \section{Resilient VAE Derivation}
\end{toappendix}

\subsection{Background: Variational autoencoders for anomaly detection}\label{sec:vae}
Variational autoencoders (VAEs)~\cite{Kingma2014_VAE} are probabilistic generative models designed for efficient inference and learning. The data is modeled as $ p(x) = \e_{p(z)}{ p_\theta(x|z) } $ where $ p(z) $ is a prior over the latent representation $ z \in \mathbb{R}^K $ and $ p_\theta(x|z) $ is the \textit{decoder} model. Typically $ K < D $, so the latent space acts as an information bottleneck, thereby implicitly assuming our data has a low-dimension representation. When $ p_\theta(x|z) $ is parameterized as a deep neural network, both the marginal likelihood $ p(x) $ and posterior likelihood $ p(z|x) $ are typically intractable~\cite{Kingma2014_VAE}; thus, approximate variational inference is required, via an \textit{encoder} model $ q_\phi(z|x) $. The resulting variational bound on the marginal log-likelihood, called the evidence lower bound (ELBO), is
\begin{align}
    \log p(x) & \geq \mathcal{L}_\text{VAE}(x) = \e_{q_\phi(z|x)} { \log p_\theta(x|z) } - \kl{q_\phi(z|x)}{p(z)} , \label{eqn:vae_loss}
\end{align}
and the lower bound $ \mathcal{L}_\text{VAE}(x) $ is maximized w.r.t.\ to the parameters $ \theta, \phi $. We refer to the first term, $ \e_{q_\phi(z|x)} { \log p_\theta(x|z) } $, as the reconstruction likelihood, and the second term, $ \kl{q_\phi(z|x)}{p(z)} $, as the prior-matching penalty.

\showto{thesis}{For tractability and simplicity, several standard modeling assumptions are typically made. First, the encoder is assumed to be a multivariate Gaussian distribution $ q_\phi(z|x) = \mathcal{N} \left( z; \mu(x;\phi), \Sigma(x;\phi) \right) $, with a mean $ \mu(x;\phi) $ and diagonal covariance matrix $ \Sigma(x;\phi) $, as this allows for the use of the re-parameterization trick of~\cite{Kingma2014_VAE} to obtain better gradient estimates w.r.t.~$\phi$. Second, the latent prior $ p(z) $ is chosen to be an isotropic multivariate Gaussian $ p(z) = \mathcal{N} \left(z; 0, I \right)$, so the KL-divergence is analytically tractable. Lastly, the decoder is assumed to be feature-separable $ p_\theta(x|z) = \prod_{d=1}^n p_\theta(x_d | z) $ (i.e., the features are conditionally independent given the latent representation).}

To apply the VAE model to anomaly detection, prior work most commonly uses the reconstruction likelihood to identify anomalies, where anomalies are assumed to be poorly reconstructed~\cite{An2015_Variational}.
Furthermore, previous studies usually train the VAE model exclusively on normal data, as otherwise, the VAE might learn to reconstruct anomalous examples present in the training data. In fact, the VAE loss is weighted towards the anomalous examples, as both the reconstruction likelihood and prior-matching penalties are analogous to minimizing an $L_2$ norm, which is not robust to outliers.
\showto{thesis}{(We show this more explicitly in~\cref{sec:vae_norobust}.)}

\begin{toappendix}
    \begin{Subsection}{thesis}{\label{sec:vae_norobust}Non-robustness of VAE loss}
    The standard VAE loss is not robust to anomalies; in fact, it is weighted towards the worst reconstructed samples (i.e., the anomalous ones!).

    To illustrate this for the reconstruction likelihood term, suppose the features are modeled by a Gaussian distribution $ p_\theta(x_d | z) = \mathcal{N} \left(x_d; \mu_d(z; \theta), \sigma_d(z; \theta)^2 \right) $. Then, the reconstruction likelihood, which we maximize, is
    \begin{align*}
        \e_{q_\phi(z|x)} { \log p_\theta(x|z) } & = \e_{q_\phi(z|x)} { \sum_d \log \left( \frac{1}{\sqrt{2 \pi} \sigma_d(z; \theta)} e^{-\frac{1}{2} \left( \frac{x_d - \mu_d(z;\theta)}{\sigma_d(z; \theta)} \right)^2 } \right) } \\
        & = -\frac{1}{2} \sum_d \e_{q_\phi(z|x)} { \left( \frac{x_d - \mu_d(z;\theta)}{\sigma_d(z; \theta)} \right)^2 } - \e_{q_\phi(z|x)} {\log \sigma_d(z; \theta)} - \frac{D}{2} \log (2 \pi).
    \end{align*}
    Maximizing this term is effectively equivalent to minimizing the $L_2$ norm of the reconstruction $z$-score, which is not robust to outliers. 
    
    To illustrate this for the prior-matching term, suppose the latent prior is $ \mathcal{N}\left(0, I\right)$ and the encoder is $ q_\phi(z|x) = \mathcal{N}\left(\mu(x;\phi), \Sigma(x;\phi) \right) $, where $ \Sigma = \text{diag}\left(\sigma_1^2, \dots, \sigma_K^2\right)$. Then, the prior-matching term, which we minimize, is
    \begin{align*}
        \kl{q_\phi(z|x)}{p(z)} & = \frac{1}{2} \left[ \norm{\mu(z;\phi)}^2 + \text{trace}(\Sigma(x;\phi)) - \log \left|\Sigma(x;\phi)\right| - D \right] \\
        & = \sum_k \left[ \frac{1}{2} \mu_k(z;\phi)^2 + \frac{1}{2} \sigma_k^2 - \log \sigma_k \right] - K
    \end{align*}
    Minimizing this term is again akin to minimizing a $L_2$ norm and is therefore not robust.
    \end{Subsection}
\end{toappendix}

\subsection{Mixture model}\label{sec:mixture_model}
Since the standard VAE is not robust to anomalies and we assume our training data is contaminated by anomalies, we propose an extension of the VAE model, which we call Resilient VAE (ResVAE), that aims to distinguish between normal and anomalous samples during training using a mixture model approach. \showto{thesis}{Our method is a generalization of RVAE~\cite{Eduardo20_RVAE}, which formed the basis of our method (we enumerate the differences in~\cref{sec:rvae_compare}).}

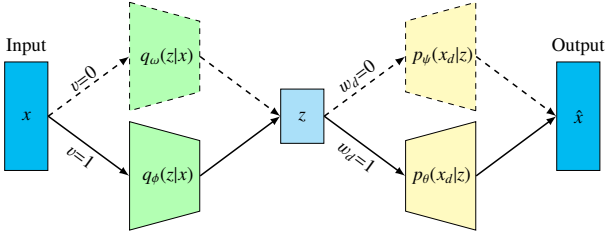
\begin{figure}[htbp]
    \centering
    \showto{chep}{\sidecaption}
    \resizebox{0.63\textwidth}{!}{
        \begin{tikzpicture}
            \tikzstyle{gen}=[-latex, thick]
            \tikzstyle{enc}=[trapezium, trapezium angle=75, minimum width=2cm, minimum height=1.3cm, trapezium stretches=true, rotate=-90, draw=black!100]
            \tikzstyle{dec}=[trapezium, trapezium angle=75, minimum width=2cm, minimum height=1.3cm, trapezium stretches=true, rotate=90, draw=black!100]
            \tikzstyle{box}=[rectangle, draw=black!100, inner sep=2pt, minimum width=8mm, minimum height=2cm]
            \node[box, fill=cyan!80, label=above:Input] (x) {$x$};
            \node[enc, fill=green!30, right=1.5cm of x, anchor=south, xshift=1.1cm, label={[align=center,anchor=east]$q_\phi(z|x)$}] (enc) {};
            \node[enc, fill=green!30, dashed, right=1.5cm of x, anchor=south, xshift=-1.1cm, label={[align=center,anchor=east]$q_\omega(z|x)$}] (out_enc) {};
            \node[box, fill=cyan!30, right=4.3cm of x, minimum height=1cm] (z) {$z$};
            \node[dec, fill=yellow!30, right=1.5cm of z, anchor=north, xshift=-1.1cm, label={[align=center,anchor=west]$p_\theta(x_d|z)$}] (dec) {};
            \node[dec, fill=yellow!30, dashed, right=1.5cm of z, anchor=north, xshift=1.1cm, label={[align=center,anchor=west]$p_\psi(x_d|z)$}] (out_dec) {};
            \node[box, fill=cyan!80, right=4.3cm of z, label=above:Output] (xhat) {$\hat{x}$};
            \draw[gen][postaction={decoration={text along path, text align=center, text={$v=1$}, raise=-10pt}, decorate}] (x.east) to (enc.south);
            \draw[gen,dashed][postaction={decoration={text along path, text align=center, text={$v=0$}, raise=2pt}, decorate}] (x.east) to (out_enc.south);
            \draw[gen][postaction={decoration={text along path, text align=center, text={${w_d}=1$}, raise=-10pt}, decorate}] (z.east) to (dec.north);
            \draw[gen,dashed][postaction={decoration={text along path, text align=center, text={${w_d}=0$}, raise=4pt}, decorate}] (z.east) to (out_dec.north);
            \path (enc.north) edge [gen] (z.west)
                (out_enc.north) edge [gen, dashed] (z.west)
                (dec.south) edge [gen] (xhat.west)
                (out_dec.south) edge [gen, dashed] (xhat.west);
        \end{tikzpicture}
    }
    \caption{Network diagram for our proposed ResVAE. ResVAE adds a sample mixing variable $ v $ and feature mixing variables $ w_d $ and splits the encoder and decoders into an inlier (solid path) and outlier (dotted path) version.}\label{fig:ResVAE_network}
\end{figure}

We begin by adding several binary latent variables to the standard VAE: a sample mixing variable $ v \in \{0,1\} $ and feature mixing variables $ w_d \in \{0,1\} $, where $1$ means inlier and 0 outlier. These mixing variables control whether an inlier or outlier path through the ResVAE model is taken, as shown in~\cref{fig:ResVAE_network}. With the new latent variables, our data is now modeled as $ p(x) = \e_{p(v,w,z)} { p(x|v,w,z) } $; we use the variational posterior $ q(v,w,z|x) $ for inference.
Our model supposes Bernoulli distributions for the mixing variables, a mixture prior on the $ z $ latent variable, and encoder and decoder mixture models:
\begin{alignat}{3}
    p(v) & = \Bernoulli{v; \rho}, & \quad q_\gamma(v|x) & = \Bernoulli{v; \gamma(x)}, \label{eqn:geninf_v} \\
    p(w_d|v) & = \Bernoulli{w_d; \alpha_d}^v \Bernoulli{w_d; 0}^{1-v}\!, & \quad q_\pi(w_d|v,x) & = \Bernoulli{w_d; \pi_d(x)}^v \Bernoulli{w_d; 0}^{1-v}\!, \label{eqn:geninf_wd_mix} \\
    p(z|v) & = p(z)^v p_o(z)^{1-v}, & \quad q(z|v,x) & = q_\phi(z|x)^v q_\omega(z|x)^{1-v}, \label{eqn:geninf_z_mix} \\
    p(x|v,w,z) & = \prod_{d=1}^D p_\theta(x_d|z)^{v w_d} p_\psi(x_d|z)^{1 - v w_d} , \label{eqn:gen_x_mix} 
\end{alignat}
where $ \rho, \alpha_d \in [0,1] $ can be used to capture prior knowledge on the degree of contamination in the training data. We revisit these prior parameters in~\cref{sec:learnable_logistic}. The functions $ \gamma(x) $ and $ \pi_d(x) $ yield estimates of the mixing variables $ v $ and $ w_d $. As we show in~\cref{sec:resvae_loss}, they have simple closed-form expressions as functions of the encoder and decoder parameters; this avoids the need for $\gamma$ and $\pi_d$ to be parameterized as separate neural networks.

\showto{thesis}{The generative model corresponds to the following process. First, a coin is flipped to determine if the entire sample is inlier ($v=1$) or outlier ($v=0$). Then, a coin is flipped for each feature to determine if that feature is inlier ($w_d=1$) or outlier ($w_d=0$), where we assume inlier samples can have outlier features but outlier samples are entirely composed of outlier features. Then, the latent representation $ z $ and data $ x $ are generated from the appropriate component of the mixture models in~\cref{eqn:geninf_z_mix,eqn:gen_x_mix}.}

Our mixture model introduced several outlier distributions---$ p_o(z), p_\psi(x|z)$, and $ q_\omega(z|x) $---whose purpose is to provide an alternate model for outlier data, thereby allowing the inlier encoder $ q_\phi(z|x) $ and decoder $ p_\theta(x|z) $ to focus on representing only the inlier data in our training data. Concretely, we define the inlier and outlier distributions as
\begin{alignat*}{5}
     p(z) & = \mathcal{N}\left(z; 0, I\right), & \quad p_\theta(x_d|z) & = p\left(x_d; \eta_d(z; \theta) \right), & \quad q_\phi(z|x) &= \mathcal{N} \big(z; \mu(x; \phi), \Sigma(x; \phi) \big), \\
     p_o(z) & = \mathcal{N}\left(z; 0, \delta_z^2 I\right), & \quad p_\psi(x_d|z) & = p\left(x_d; \eta_d(z; \theta_0) / \delta_x^2 \right), & \quad q_\omega(z|x) & = q_{\phi_0}(z|x),
\end{alignat*}
where $ \delta_x, \delta_z $ are hyperparameters of the method, $ p_\theta(x_d|z) $ is taken to be an exponential family distribution with natural parameters $ \eta_d(z; \theta) $, $ \theta_0 $ is a copy of $ \theta $, and $ \phi_0$ is a copy of $ \phi $ (where the copy prevents gradients via automatic differentiation).
(We explain these modeling choices in~\cref{sec:outlier_dist_choices}.)

\begin{toappendix}
    \subsection{Choice of outlier distributions}\label{sec:outlier_dist_choices}
    We might be tempted to parameterize the outlier encoder $ q_\omega(z|x) $ and decoder $ p_\psi(x|z) $ as independent neural networks and jointly train the two VAEs to represent the data, with the outlier VAE representing the outlier data. However, this would require assuming that the outlier data has a low-dimension representation. Since we expect anomalies to be varied and largely originate from low-density regions, we want to avoid making such an assumption, and therefore we make a different choice.

    We choose to model both the outlier latent prior $ p_o(z) $ and the outlier decoder $ p_\psi(x|z) $ as more ``diffuse'' versions of their inlier counterparts, where the ``diffuse'' versions should better model outliers than the inlier models. Formally, for any exponential family distribution $ p(y) = h(y) e^{\eta^T T(y) - A(\eta)} $ with natural parameters $ \eta $, we define its $\delta$-diffuse ($\delta > 1$) counterpart as having scaled the natural parameters by $ 1 / \delta^2 $, that is $ \eta \rightarrow \eta/\delta^2$. \showto{thesis}{(We describe this transformation in greater detail in~\cref{sec:diffuse_exp_outlier}.)} The hyperparameters $ \delta_x, \delta_z $ control how ``diffuse'' the outlier models are relative to the inlier models. Also, for $ p_\psi(x_d|z)$, we crucially use a copy $\theta_0$ of the current $ \theta $ to prevent gradients to $ \theta $ through $ p_\psi(x_d|z) $; that is, we ensure inlier decoder weights are not updated through outlier decoder.

    For the outlier encoder $ q_\omega(z|x) $, we might be tempted to similarly set it to be a ``diffuse'' version of $ q_\phi(z|x) $. However, this would require an extra traversal through the inlier decoder $ p_\theta(x|z) $ (via $ p_\psi(x|z) $) since outliers are reconstructed through $ \e_{q_\omega(z|x)}{p_\psi(z|x)} $ (as shown in~\cref{eqn:full_loss}). But this would assess the inlier decoder over outlier regions of the latent space that it is not trained on. Therefore, we set $ q_\omega(z|x) = q_{\phi_0}(z|x) $ (as in~\cite{Daniel19_VarSAD}), where we again use a copy $ \phi_0 $ of $ \phi$ to prevent gradient flow.

    \showto{thesis}{
    \subsection{``Diffuse'' exponential family distributions}\label{sec:diffuse_exp_outlier}
    Consider an exponential family distribution $ p(y; \eta) = h(y) e^{\eta^T T(y) - A(\eta)} $, where $ \eta $ are the natural parameters, $ T(x) $ is the sufficient statistic, $ A(\eta) $ is the log-partition function, and $ h(x) $ is a normalization factor. Let us consider the transformation $ \eta \rightarrow \eta/\delta^2$ for several common exponential family distributions. We will show the transformation w.r.t.~the canonical parameters for easier interpretation.
    \begin{enumerate}
        \item \textit{Bernoulli/Continuous Bernoulli}: $ p \rightarrow \sigmoid{\logit{p} / \delta^2} $. Note that this tends towards the uniform distribution $ U(0,1) $ as $ \delta \rightarrow \infty $. Equivalent to temperature scaling ~\cite{Guo17_Cali,Hinton15_Distill}.
        \item \textit{Beta}: $ [\alpha, \beta] \rightarrow \left[1 + \frac{\alpha - 1}{\delta^2}, 1 + \frac{\beta - 1}{\delta^2} \right] $. Again note that this tends towards the uniform distribution $ U(0,1) $ as $ \delta \rightarrow \infty $.
        \item \textit{Categorical/Multinomial}: $ \left[p_1, p_2, \dots, p_k \right] \rightarrow \left[p_1^{1/\delta^2}, \dots, p_k^{1/\delta^2} \right] $, up to a probability normalization to $ 1 $. Note that this tends towards the uniform distribution over all $ k $ categories as $ \delta \rightarrow \infty $. Equivalent to using a power heuristic in importance sampling or tempering the distribution~\cite{Owen00_Import,Kapoor22_Uncertainty,Neal96_Sampling}.
        \item \textit{Gaussian}: $ \left[ \mu, \Sigma \right] \rightarrow \left[ \mu, \frac{1}{\delta^2} \Sigma \right] $. Note that this inflates the standard deviation by $ \delta $ and therefore creates broader Gaussians as $ \delta $ increases. Previously used in~\cite{Quinn09,Gales99_TailDist} for modelling heavier tails.
    \end{enumerate}
    }
    
\end{toappendix}

\subsection{ResVAE loss and coordinate ascent}\label{sec:resvae_loss}
Instead of maximizing the ELBO, we adopt the constrained-optimization approach of $\beta$-VAE~\cite{Higgins17_BetaVAE}, which introduces several coefficients regularizing the prior-matching penalties. Our objective is then to maximize the following loss function:
\begin{align*}
    \max_{\phi, \theta, \omega, \psi} \mathcal{L}(x) & = \e_{q(v,w|x)}{ \e_{q(z|v,x)}{\log p(x|v,w,z)} } - \beta_1 \e_{q_\gamma(v|x)}{ \kl{q(z|x,v)}{p(z|v)}} \\
    & \quad - \sum_{d=1}^D \frac{1}{\beta_{2,d}} \e_{q_\gamma(v|x)}{\kl{q_{\pi_d}(w_d|v,x)}{p(w_d|v)}} - \frac{1}{\beta_3} \kl{q_\gamma(v|x)}{p(v)}. \nonumber
\end{align*}
(A derivation of this loss can be found in~\cref{sec:lagrange_form}.) $ \beta_1 $ plays an identical role to that in $\beta$-VAE~\cite{Higgins17_BetaVAE}, controlling the latent information bottleneck against the reconstruction objective. We discuss the other regularization coefficients $ \beta_{2,d}, \beta_3 $ in~\cref{sec:learnable_logistic}.

\begin{toappendix}
    \subsection{Lagrangian formulation}\label{sec:lagrange_form}
    Following the framework of $\beta$-VAE~\cite{Higgins17_BetaVAE}, we define the ResVAE objective as solving the following optimization problem:
    \begin{alignat}{1}
        \max \quad & \e_{q(v,w,z|x)} {\log p(x|v,w,z) } \\
        \text{s.t.} \quad & \e_{q_\gamma(v|x)} { \kl{q(z|v,x)}{p(z|v)} } < \epsilon_1 \nonumber \\
        & \e_{q_\gamma(v|x)} { \kl{q_\pi(w_d|v,x)}{p(w_d|v)} } < \epsilon_{2,d} \nonumber\\
        & \kl{q_\gamma(v|x)}{p(v)} < \epsilon_3. \nonumber
    \end{alignat}
    By the KKT conditions~\cite{Kuhn51_KKT,Karush39}, we can write this as the Lagrangian:
    \begin{align}
        \mathcal{F}(x) = \ & \e_{q(v,w,z|x)} {\log p(x|v,w,z) } \\
        & - \beta_1 \left( \e_{q_\gamma(v|x)} { \kl{q(z|v,x)}{p(z|v)} } - \epsilon_1 \right) \nonumber \\
        & - \sum_d \frac{1}{\beta_{2,d}} \left( \e_{q_\gamma(v|x)} { \kl{q_\pi(w_d|v,x)}{p(w_d|v)} } - \epsilon_{2,d} \right) \nonumber \\
        & - \frac{1}{\beta_3} \left( \kl{q_\gamma(v|x)}{p(v)} - \epsilon_3 \right). \nonumber
    \end{align}
    We have $ \beta_1, \beta_{2,d}, \beta_3 \geq 0 $ by complementary slackness. The latter two Lagrange multipliers are written as reciprocals for easier interpretation of their role, as we'll see shortly. Dropping the constants, we recover the loss function:
    \begin{align}
        \mathcal{L}(x) = \ & \e_{q(v,w|x)}{ \e_{q(z|v,x)}{\log p(x|v,w,z)} } \\
        & - \beta_1 \e_{q_\gamma(v|x)}{ \kl{q(z|x,v)}{p(z|v)}} \nonumber \\
        & - \sum_d \frac{1}{\beta_{2,d}} \e_{q_\gamma(v|x)}{\kl{q_{\pi_d}(w_d|v,x)}{p(w_d|v)}} \nonumber \\
        & - \frac{1}{\beta_3} \kl{q_\gamma(v|x)}{p(v)}. \nonumber
    \end{align}
    Using the generative and inference models described in~\cref{eqn:geninf_v,eqn:geninf_wd_mix,eqn:geninf_z_mix,eqn:gen_x_mix}, our loss is concretely
    \begin{align}
        \mathcal{L}(x) = \ & \gamma(x) \sum_d \pi_d(x) \e_{q_\phi(z|x)} {\log p_\theta(x_d|z) } \label{eqn:full_loss} \\
        & + \gamma(x) \sum_d \left(1 - \pi_d(x) \right) \e_{q_\phi(z|x)} {\log p_\psi(x_d|z) } \nonumber \\
        & + \left(1 - \gamma(x) \right) \sum_d \e_{q_\omega(z|x)} {\log p_\psi(x|z) } \nonumber \\
        & - \beta_1 \gamma(x) \kl{q_\phi(z|x)}{p(z)} \nonumber \\
        & - \beta_1 \left(1 - \gamma(x) \right) \kl{q_\omega(z|x)}{p_o(z)} \nonumber \\
        & - \gamma(x) \sum_d \frac{1}{\beta_{2,d}} \kl{\Bernoulli{\pi_d(x)}}{\Bernoulli{\alpha_d}} \nonumber \\
        & - \frac{1}{\beta_3} \kl{\Bernoulli{\gamma(x)}}{\Bernoulli{\rho}}. \nonumber
    \end{align}
\end{toappendix}

We can simplify this loss function by performing coordinate ascent (as in~\cite{Jordan99_VarMethods,Eduardo20_RVAE}) on $ \gamma(x) $ and $ \pi(x) $ since $\mathcal{L}(x)$ is convex in these variables for each $ x $. Their optimal values are:
\begin{equation}
    \pi_d(x) = \sigmoid{\beta_{2,d} r_d(x) + \logit{\alpha_d}}, \qquad \gamma(x) = \sigmoid{\beta_3 g(x) + \logit{\rho}}, \label{eqn:pi_gamma_opt}
\end{equation}
where $ \sigmoid{\cdot} $ is the sigmoid function, $ \logit{\cdot} $ is the logit function, and $ r_d(x) $ and $ g(x) $ can be interpreted as likelihood ratio tests~\cite{VanTrees04}
(i.e., $ r_d(x) > 0 $ suggests $x_d$ is drawn from the inlier model instead of the outlier model and the opposite for $ r_d(x) < 0 $).
\showto{thesis}{Therefore, $ \pi_d(x) $ and $ \gamma(x) $ should be closer to $ 1 $ for inliers and $ 0 $ for outliers.}
(A derivation \showto{thesis}{of this coordinate-ascent optimal solution }and the definition of $ r_d(x) $ and $ g(x) $ can be found in~\cref{sec:coord_ascent}.)

\begin{toappendix}
    \subsection{Coordinate ascent}\label{sec:coord_ascent}
    We first state two identities for the KL divergence of two Bernoulli distributions (these can be readily shown):
    \begin{alignat}{1}
        \frac{\partial}{\partial a} \kl{\Bernoulli{a}}{\Bernoulli{b}} & = \logit{a} - \logit{b}, \\
        \frac{\partial^2}{\partial a^2} \kl{\Bernoulli{a}}{\Bernoulli{b}} & = \frac{1}{a (1-a)} > 0,
    \end{alignat}
    where $ \logit{x} = \log \frac{x}{1-x} $ is the logit function.
    
    Now, since $ \mathcal{L}(x) $ is concave in $ \pi_d(x) $, we can directly optimize for it, holding all other variables fixed. Taking the first two derivatives, and using the above identities, yields:
    \begin{align*}
        \frac{\partial}{\partial \pi_d(x)} \mathcal{L}(x) & = \gamma(x) \left[ r_d(x) - \frac{\logit{\pi_d(x)} - \logit{\alpha_d}}{\beta_{2,d}} \right], \\
        \frac{\partial^2}{\partial \pi_d(x)^2} \mathcal{L}(x) & = - \frac{1}{\beta_{2,d}} \frac{1}{\pi_d(x) \left( 1-\pi_d(x) \right)} < 0, \\
        \frac{\partial^2}{\partial \pi_d(x) \partial \pi_j(x)} \mathcal{L}(x) & = 0, \; \text{ if } j \neq d,
    \end{align*}
    where $ r_d(x) = \e_{q_\phi(z|x)} {\log \frac{p_\theta(x_d|z)}{p_\psi(x_d|z) } } $. The optimal $ \pi_d(x) $ can therefore be written:
    \begin{equation}
        \pi_d(x) = \sigmoid{\beta_{2,d} r_d(x) + \logit{\alpha_d}},
    \end{equation}
    where $ \sigmoid{x} = \left( 1 + e^{-x} \right)^{-1} $ is the sigmoid function.
    
    Similarly, $ \mathcal{L}(x) $ is concave in $ \gamma(x) $, so we can directly optimize for it. The first two derivatives are:
    \begin{align*}
        \frac{\partial}{\partial \gamma(x)} \mathcal{L}(x) & = g(x) - \frac{\logit{\gamma(x)} - \logit{\rho}}{\beta_3}, \\
        \frac{\partial^2}{\partial \gamma(x)^2} \mathcal{L}(x) & = - \frac{1}{\beta_3} \frac{1}{\gamma(x) \left( 1-\gamma(x) \right)} < 0,
    \end{align*}
    where
    \begin{align*}
        g(x) & = \beta_1 s(x) + \sum_d \left( \pi_d(x) r_d(x) - \frac{1}{\beta_{2,d}} \kl{\Bernoulli{\pi_d(x)}}{\Bernoulli{\alpha_d}} \right), \\
        s(x) & = \kl{q_\omega(z|x)}{p_o(z)} - \kl{q_\phi(z|x)}{p(z)}.
    \end{align*}
    Using the fact that $ q_\omega(z|x) = q_{\phi_0}(z|x) $, we get a simpler form: $ s(x) = \e_{q_\phi(z|x)}{\log \frac{p(z)}{p_o(z)}} $.
    We can also simplify $ g(x) $ with the following identity. For $ y = \sigmoid{a x + \logit{b}} $,
    \[
        y x - \frac{1}{a} \kl{\Bernoulli{y}}{\Bernoulli{b}} = \frac{1}{a} \log \frac{1-b}{1-y}.
    \]
    Thus we can write:
    \[
        g(x) = \beta_1 s(x) + \sum_d \frac{1}{\beta_{2,d}} \log \frac{1 - \alpha_d}{1 - \pi_d(x)} .
    \]
    Lastly, we quickly note that the gradient of the full loss (\cref{eqn:full_loss}) w.r.t.~$\phi$ and $ \theta $ can be calculated without backpropagation through $ \pi_d(x) $ and $ \gamma(x) $. This follows directly from the chain rule, where
    \[
        \frac{\partial \mathcal{L}(x)}{\partial \phi} = \sum_d \frac{\partial \mathcal{L}(x)}{\partial \pi_d(x)} \frac{\partial \pi_d(x)}{\partial \phi} + \frac{\partial \mathcal{L}(x)}{\partial \gamma(x)} \frac{\partial \gamma(x)}{\partial \phi} + \dots,
    \]
    since $ \frac{\partial \mathcal{L}(x)}{\partial \pi_d(x)} = \frac{\partial \mathcal{L}(x)}{\partial \gamma(x)} = 0 $ by coordinate ascent.
    
\end{toappendix}

Using the closed form for $\pi_d(x) $ and $\gamma(x)$ and the given outlier distributions, and removing untrainable terms, our ResVAE loss function becomes:
\begin{equation}
    \max_{\phi, \theta} \mathcal{L}_\text{ResVAE}(x) = \gamma(x) 
    \bigg[ \sum_d \pi_d(x) \e_{q_\phi(z|x)} {\log p_\theta(x_d|z) } - \beta_1 \kl{q_\phi(z|x)}{p(z)} \bigg]. \label{eqn:resvae_loss}
\end{equation}
This is nearly the form of the standard VAE loss (or $\beta$-VAE loss~\cite{Higgins17_BetaVAE}), with the addition of several weighting factors that either preserve the sample (or features) in the loss function or remove it. $\gamma(x) $ weights entire samples and $ \pi_d(x) $ weights individually features.
\showto{thesis}{(We note that the coordinate ascent approach implies the backward pass can be simplified to avoid gradients through $ \pi_d(x) $ and $ \gamma(x) $ as shown in~\cref{sec:coord_ascent}).}

\subsection{Learnable logistic parameters} \label{sec:learnable_logistic}
We can interpret the form of $ \pi_d(x) $ and $ \gamma(x) $ as a logistic regression, where the input variables are the likelihood ratios $ r_d(x) $ and $ g(x) $ respectively and the parameters are the regularization coefficients $ \beta_{2,d}, \beta_3 $ and contamination prior parameters $ \alpha_d, \rho $. Ideally, $ \pi_d(x) $ and $ \gamma(x) $ should be nearly $ 1 $ for inliers and $ 0 $ for outliers. Achieving this requires setting the logistic parameters appropriately.
\showto{thesis}{(We demonstrate this difficulty in~\cref{sec:logistic_difficulty}).}

\begin{toappendix}
    \begin{Subsection}{thesis}{\label{sec:logistic_difficulty}Logistic calibration difficulty}
    To illustrate the difficulties in setting the logistic parameters a priori, we focus on the $ \pi_d(x) $ logistic regression for clarity, noting that the same type of analysis applies to $ \gamma(x) $ as well. We again suppose the features are modeled by a Gaussian distribution $ p_\theta(x_d | z) = \mathcal{N} \left(x_d; \mu_d(z;\theta), \sigma_d(z; \theta)^2 \right) $ and $ p_\psi(x_d|z) $ is a $\delta_x$-diffuse version. Then,
    \begin{align}
        \pi_d(x) & = \sigmoid{ - \frac{\beta_{2,d}}{2} \left(1 - \frac{1}{\delta_x^2}\right) z_\text{score}^2 + \log \delta_x + \logit{\alpha_d} } ,
    \end{align}
    where $ z_\text{score} = \e_{q_\phi(z|x)} {\left( \frac{x_d - \mu_d(z;\theta)}{\sigma_d(z; \theta)} \right)^2} $ is the expected z-score (i.e., expected number of standard deviations from the mean). To simplify our discussion, note that, for any $ \delta_x $, $ (\beta_{2,d}, \alpha_d) $ parameterize all possible scale and shifts for the logistic regression, so we fix $ \delta_x = 2 $. 
    
    We perhaps are inclined to simply set $ \beta_{2,d} = 1 $ (as in the ELBO formulation) and $ \alpha_{2,d} $ to match our prior knowledge of the contamination of our dataset (say $ \alpha_{2,d} = 0.95 $ as in RVAE~\cite{Eduardo20_RVAE}). The resulting plot of $ \pi_d(x) $ against $ z_\text{score} $ is shown in~\cref{fig:pi_plot}. For a perfect reconstruction $ z_\text{score} = 0 $ (which we expect only inliers to achieve), $ \pi_d(x) \approx 0.97 $, so inliers do achieve $ \pi_d(x) $ near $ 1 $. However, for a poor reconstruction $ z_\text{score} = 3 $ (which we expect to indicate an outlier as only $ 0.25\%$ of samples from $\mathcal{N}\left(0,I\right)$ have a worse z-score), $ \pi_d(x) \approx 0.57 \gg 0 $. Clearly $ \pi_d(x) $ under this setting is doing a poor job of dropping outlier data from our loss function. We could instead set $ \beta_{2,d} $ to be a higher value (say $ \beta_{2,d} = 2 $), so that $ \pi_d(x) \approx 0 $ for poor reconstructions. However, this creates a problem at the beginning of training. At the beginning of training, we do not generally expect the data to be well-reconstructed; therefore, much of the data might lie in the saturating region near $ \pi_d(x) = 0 $. But this implies the weights are nearly identical for all inputs, whether inliers or not, whether $ z_\text{score} = 3 $ or $ z_\text{score} = 10 $.
    
    \begin{figure}[htbp]
        \centering
        \showto{chep}{\sidecaption}
        \includegraphics[width=0.6\textwidth,clip]{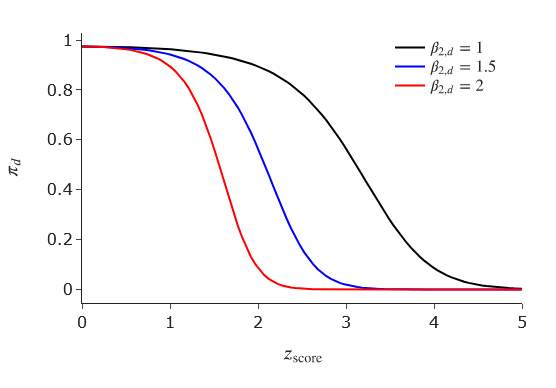}
        \caption{Feature weight $ \pi_d(x) $ for Gaussian-modeled data with reconstruction z-score $ z_\text{score} $ for various values of $ \beta_{2,d} $, under the setting $ \alpha_d=0.95$, and $ \delta_x = 2$.}
        \label{fig:pi_plot}
    \end{figure}
    \end{Subsection}
\end{toappendix}

To ensure at all stages of training that $ \pi_d(x) $ and $ \gamma(x) $ are expressive, we propose to learn the logistic parameters. Since we can interpret $ \pi_d(x) $ and $ \gamma(x) $ as discriminators (as in generative adversarial nets~\cite{Goodfellow14_GAN}), we train them as binary classifiers with the binary cross entropy loss:
\begin{equation*}
    \max_{\beta_{2,d}, \alpha_d} \mathcal{L}_{\pi_d}(x) =  \e_{p(w_d|v=1,x)}{ \log q_\pi(w_d|v=1,x) }, \qquad \max_{\beta_3, \rho} \mathcal{L}_{\gamma}(x) = \e_{p(v|x)}{ \log q_\gamma(v|x) },
\end{equation*}
where $ q_\pi(w_d|v=1,x) $ depends on the logistic parameters $ \beta_{2,d}, \alpha_d $ through the definition of $ \pi_d(x) $ and $ q_\gamma(v|x) $ similarly depends on $ \beta_3, \rho $ through $ \gamma(x) $. \showto{thesis}{This formulation, in fact, is equivalent to the framing of Adversary Activated VAEs~\cite{Hu2018_Unifying} under certain assumptions. (We explore this connection further in~\cref{sec:connect_to_aavae}.) }The label distributions of the binary cross entropy terms are the posterior distributions $ p(w_d|v=1,x) = \Bernoulli{w_d; \tilde{w}_d(x)} $ and $ p(v|x) = \Bernoulli{v; \tilde{v}(x)} $. However, since these posteriors are intractable (which originally motivated the use of approximate posteriors), the labels $ \tilde{w}_d(x) $ and $ \tilde{v}(x) $ must be approximated.

In defining our label approximations, we want to achieve two goals. First, we want the empirical distributions of $ \pi_d(x) $ and $ \gamma(x) $ to match our prior knowledge of the dataset's contamination. Since $\rho, \alpha_d$ are now learned as logistic parameters instead of specified as contamination priors, we re-introduce the idea of a contamination prior through new prior label distributions $ f_{w_d}(w_d) $ and $ f_v(v) $ over the entire dataset. Second, since we assume that inlier examples have higher likelihood ratios $ r_d(x) $ and $ g(x) $ than outlier examples, we want $ \pi_d(x), \gamma(x) $ to be increasing in the likelihood ratios $ r_d(x), g(x) $ respectively. Therefore, we approximate the labels as:
\begin{equation}
    \tilde{w}_d(x) \approx F_{w_d} \left( \mathbb{P}_{y \sim \mathcal{D}}\left( \pi_d(x) \geq \pi_d(y) \right)\right), \qquad \tilde{v}(x) \approx F_{v}\left(\mathbb{P}_{y \sim \mathcal{D}}\left( \gamma(x) \geq \gamma(y) \right)\right), \label{eqn:pi_gamma_bce}
\end{equation}
where the inner rank statistic terms ensure the property of increasing in likelihood ratio and the cumulative distribution functions (CDFs) $ F_{w_d}(w_d) $ and $ F_v(v) $ (of $ f_{w_d}(w_d) $ and $ f_v(v) $ respectively) ensure the prior-empirical label distribution matching.

\begin{toappendix}
    \begin{Subsection}{thesis}{Connection to adversary activated VAEs}\label{sec:connect_to_aavae}
    Adversary activated VAEs (AAVAEs)~\cite{Hu2018_Unifying} connect the VAE framework with GANs~\cite{Goodfellow14_GAN} by replacing the (implicit) perfect discriminator in VAEs with a discriminator network to get an adapted VAE objective:
    \begin{align}
        \max_{\phi,\theta} \mathcal{L}_{\phi,\theta} = \e_{p_{\theta_0}(x)} { \e_{q_\phi(z|x,y) q^r_\eta(y|x)} {\log p_\theta(x|z,y)} - D_{KL}\left( q_\phi(z|x,y) q^r_\eta(y|x) || p(z|y) p(y) \right)  },
    \end{align}
    where $ p_{\theta_0}(x) = \e_{p(y)} { p_{\theta_0}(x|y) } $ is the data ``prior'' (mixing real and generated data), $ q^r_\eta(y|x) = q_\eta(1 - y | x) $ is the reversed discriminator, and
    \begin{align}
        p_\theta(x|z,y) & = \begin{cases} p_\theta(x|z), \quad \text{if $y=0$,} \\ p_\text{data}(x), \quad \text{otherwise}. \end{cases}
    \end{align}
    The discriminator is then trained as a classifier:
    \begin{align}
        \max_\gamma \mathcal{L}_\gamma = \e_{p_\theta(x|z,y) p(z|y) p(y)} { \log q_\gamma(v|x) }.
    \end{align}
    This forces the VAE to focus only on examples that appear real ($ q^r_\eta(y=1|x) \approx 1 $).
    
    We can, in fact, recover our approach through a particular setting of AAVAEs. We replace the label $ y $ with the pair $ (v,w) $, where $v$ and $w=\{w_d\}$ are our mixing variables. The VAE-training objective of AAVAE is then exactly~\cref{eqn:full_loss}, when setting the regularization coefficients to $ 1 $, taking $ q^r_\eta(v,w|x) = q^r_\pi(w|v,x) q^r_\gamma(v|x) $, and using our data model (\cref{eqn:geninf_v,eqn:geninf_wd_mix,eqn:geninf_z_mix,eqn:gen_x_mix}). The discriminator objective is also equal to~\cref{eqn:pi_gamma_bce}. However, unlike AAVAEs, which have known labels on which the discriminator can be trained, we must use the pseudo-label approach set out in~\cref{sec:learnable_logistic}.
    \end{Subsection}

    \begin{Subsection}{thesis}{Solving for logistic parameters}
    We solve for the optimal logistic parameters as a subproblem of each iteration of the ResVAE loss, using iteratively reweighted least squares (i.e., Newton's method) on an (optional) buffer of recent $ \pi_d $ and $ \gamma $ values. This helps ensure a logistic fit suitable for the entire dataset $ \mathcal{D} $ instead of just the current minibatch. The iteratively least squares solver is implemented on the GPU to concurrently solve for all $ \pi_d $ instances (of which there are $D$) simultaneously and uses backtracking line search for robustness; the same solver works for $ \gamma $ as well. Also, we only solve the subproblem every $ n = 10$ iterations and normalize the loss function at iteration by a running factor that is updated with $ \sum_{i,d} \gamma(x_i) \pi_d(x_i) $ for a batch of examples $ \{x_i\} $. This was needed to stabilize the magnitude of the loss across different iterations; without this normalization, training ResVAE with the Adam optimizer is prone to oscillations due to loss magnitude variation.
    \end{Subsection}
\end{toappendix}

\section{Application to LCLS}\label{sec:lcls_application}

We now apply our ResVAE method to detecting anomalies at LCLS. We consider pulse-by-pulse measurements of the beam, as measured by stripline beam position monitors (BPMs)~\cite{Straumann07_BPM} and recorded by the beam-synchronous acquisition system~\cite{kim_bsa}. Each of the $150$ BPMs records three signals (X, Y, and TMIT) for each electron pulse, where X and Y measure transverse position and TMIT (transmitted intensity) measures the passing beam charge, for a total of $ 450 $ signals. Since each pulse should be i.i.d.\ under normal operation, we currently consider each pulse to be a separate input; using windows of consecutive pulses is left for future work.

Since the operators of LCLS are constantly changing the machine settings, we limit our study to periods of “steady” operation at 120 Hz (as detailed in~\cite{Humble2022_RFFault}) and train on a single day of data at a time. Nonetheless, due to the contamination in the training data, the standard VAE~\cite{Kingma2014_VAE}, RVAE~\cite{Eduardo20_RVAE}, and DSVDD~\cite{Ruff18_DSVDD} methods, as well as ResVAE without logistic learning, fail to train; they experience disrupting gradients due to the anomalous examples. Gradient clipping does not fix the training failures, as this solely moderates the magnitudes of applied gradients instead of ignoring the gradient contribution of anomalous examples entirely. We train our ResVAE model using the PyTorch framework~\cite{Paszke19_pytorch} and the Adam optimizer~\cite{Kingma14_Adam} using mini-batches. (Full details on the training setting and architecture are found in~\cref{sec:lcls_exp_settings}.)

\begin{figure}[htbp]
    \showto{chep}{\vspace*{-1em}}
    \centering
    \showto{chep}{\sidecaption}
    \resizebox{0.56\textwidth}{!}{
        \begin{tikzpicture}
            \node[anchor=south west,inner sep=0] at (0,0) {\includegraphics[trim={0 0 2cm 2cm},clip]{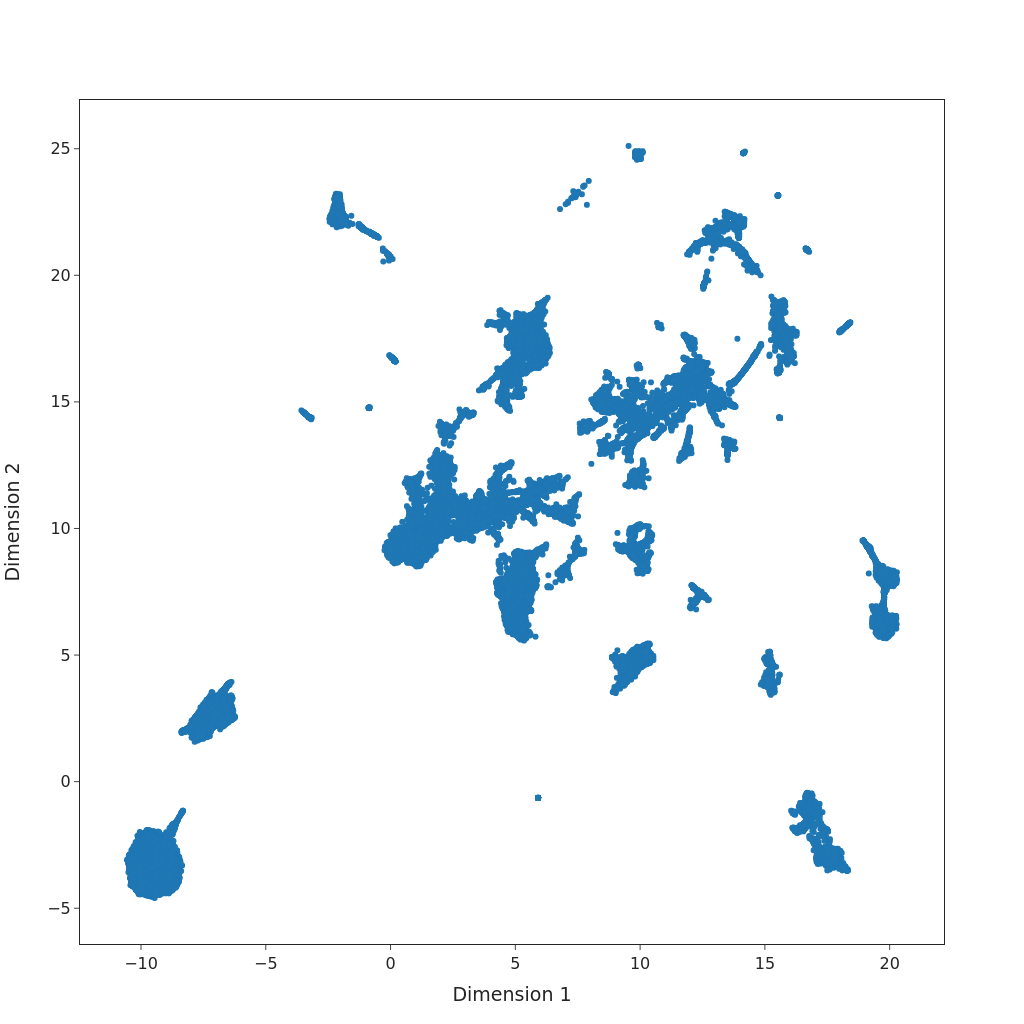}};
            \node[] (Anom1) at (24,7) {\Huge \cref{fig:x_anom}};
            \draw (Anom1.east) to (28.6, 6.5);
            \node[] (Anom2) at (10.5,5) {\Huge \cref{fig:tmit_anom}};
            \draw (Anom2.west) to (6, 6);
            \node[] (Anom3) at (14,25) {\Huge \cref{fig:beam_loss}};
            \draw (Anom3.east) to (18, 22.5);
        \end{tikzpicture}
    }
    \caption{UMAP~\cite{McInnes2018_UMAP} clustering of the feature-level anomaly attribution vector $ \pi(x) $ for anomalous pulses (where $\gamma(x) \leq 0.5 $).
    }\label{fig:lcls_anom_clustering}
    \showto{chep}{\vspace*{-1.5em}}
\end{figure}

\begin{figure}[htbp]
    \centering
    \showto{chep}{\sidecaption}
    \includegraphics[width=0.7\textwidth,trim={0 0 0 2cm},clip]{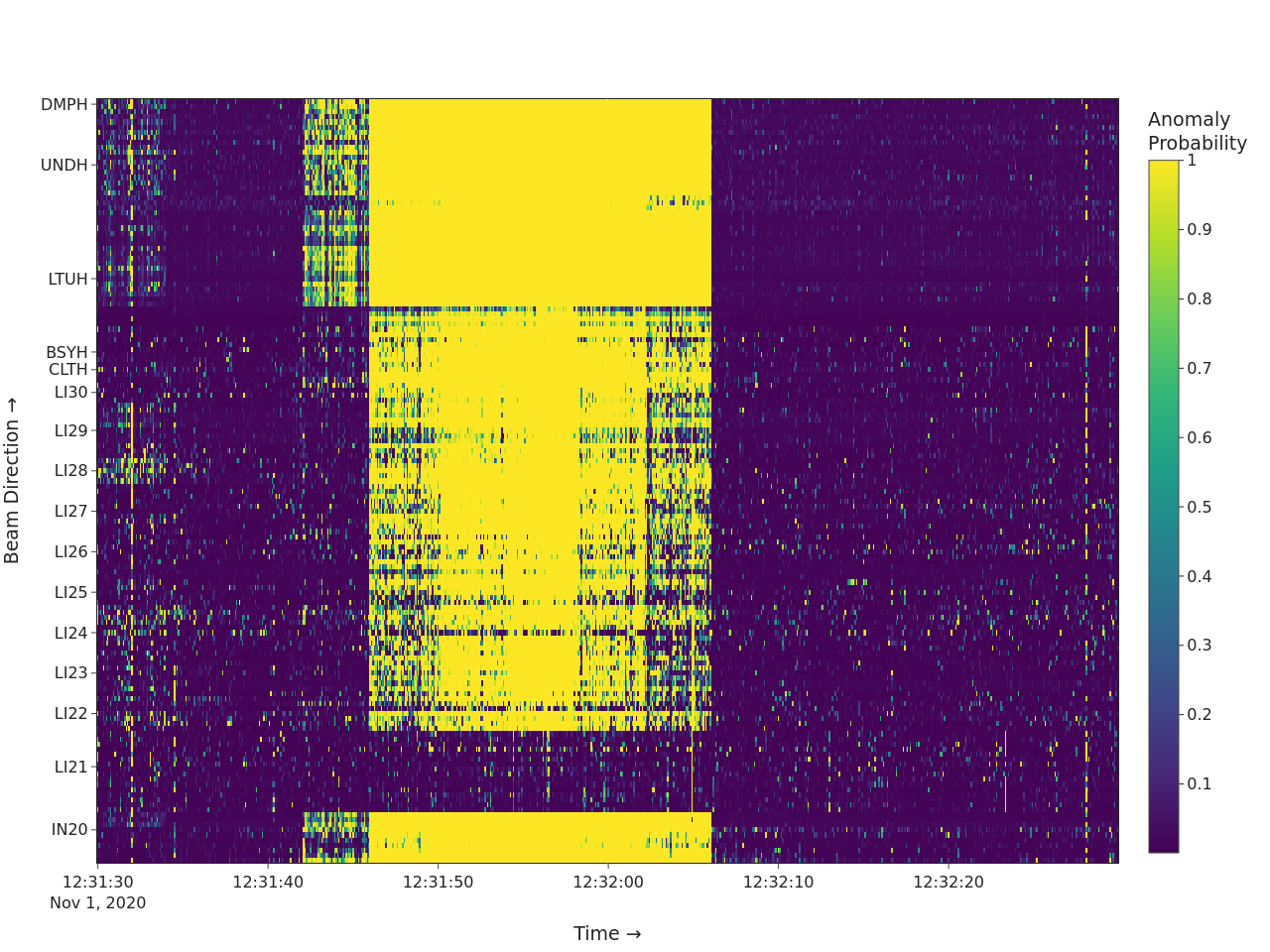}
    \caption{Example of a beam-loss anomaly detected by ResVAE and its feature-level anomaly attribution (where the feature-level anomaly probability is $ 1 - \pi_d(x) $). We only show the TMIT signals. We plot an illustrative subset of these signals in~\cref{fig:beam_loss_signal}.}\label{fig:beam_loss}
    \showto{chep}{\vspace*{-1.3em}}
\end{figure}

\begin{figure}[htbp]
    \centering
    \showto{chep}{\sidecaption}
    \includegraphics[width=0.7\textwidth,trim={0 0 0 2cm},clip]{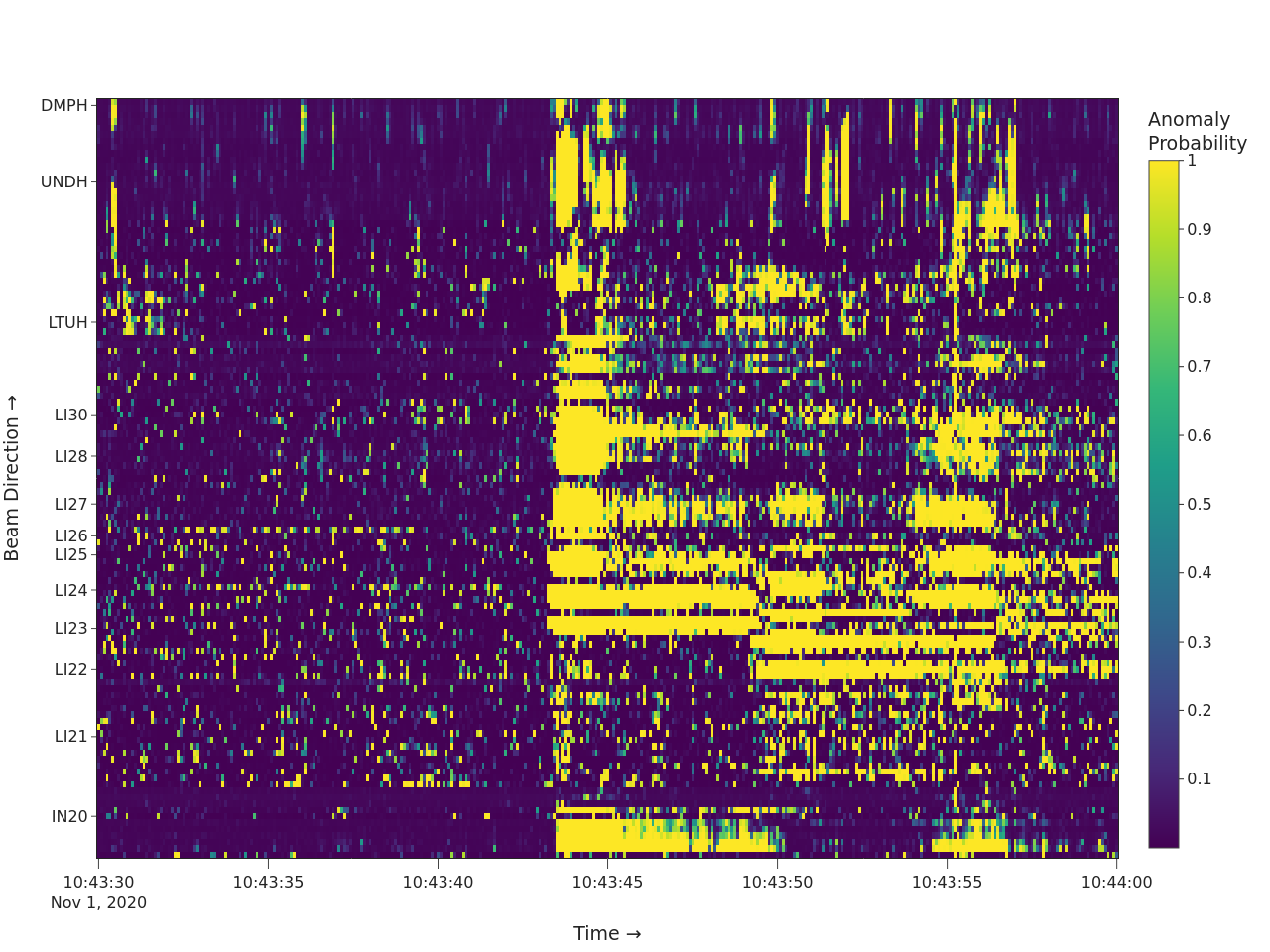}
    \caption{Example of anomaly detected by ResVAE and its feature-level anomaly attribution (where the feature-level anomaly probability is $ 1 - \pi_d(x) $). We only show the X signals. We plot an illustrative subset of these signals in~\cref{fig:x_anom_signal}.}\label{fig:x_anom}
    \showto{chep}{\vspace*{-1.5em}}
\end{figure}

To explore the anomalies detected by our method, we extract all anomalous pulses (where $ \gamma(x) \leq 0.5 $) and then perform clustering on the vector $ \pi(x) $ (which provides feature-level attribution) using UMAP~\cite{McInnes2018_UMAP}. We show the existence of several anomaly clusters in~\cref{fig:lcls_anom_clustering}. We also dive into several clusters to determine the nature of the anomaly. \cref{fig:beam_loss} shows an anomaly where the X-ray beam is partially lost, observed in the beam's TMIT signals. \cref{fig:x_anom} displays the onset of an anomaly observed in the beam's X signals. ResVAE is capable of providing operators with information regarding both the timing and location of the anomaly.


\begin{toappendix}
    \section{LCLS Experiment Details}

    \subsection{\label{sec:lcls_exp_settings}Training setup}
    We train our model on 450 real-valued signals from the stripline BPMs, where we get X, Y, and TMIT signals from 150 distinct BPMs for each electron pulse. We normalize the data using the median and MAD (median absolute deviation) statistics from periods of ``steady'' operation (see~\cite{Humble2022_RFFault}) from November 2020 through September 2022. We supplement this data with three machine setting process variables: REFS:LI21:231:EDES, REFS:LI24:790:EDES, and REFS:LI30:901:EDES. We incorporate this machine setting data using the conditional VAE architecture~\cite{Sohn15_CVAE}; that is, we condition both our encoder model and decoder model on these machine settings, by stacking the machine settings into the input to each model. We model each input feature as a Gaussian distribution.

    We set the latent dimension $ K = 64 $; latent regularization coefficient $ \beta_1 = 3.5 $; outlier hyperparameters $ \delta_z = \delta_x = 2$; Adam learning rate to $ 1e-4$; Adam weight decay to $ 1e-6 $; and the logistic learning label priors $ f_{w_d}(w_d) = \mathcal{B} \left( w_d; (1 - \nu_d) \lambda_d, \nu_d \lambda_d \right) $ and $ f_v(v) = \mathcal{B} \left( v; (1 - \mu) \kappa, \mu \kappa \right) $ with means $ \nu_d = \mu = 0.9 $ and the strengths $ \lambda_d = \kappa = 100 $. We train for at most 400 epochs with early stopping on the validation loss (stopping after 25 epochs without improvement) after splitting the data as 90\%/10\% train/validation. We use a batch size of $8192$ pulses and shuffle the data. Our encoder and decoder architectures are shown in~\cref{fig:lcls_arch}.

    \begin{figure}[htbp]
        \centering
        \subfigure[Encoder architecture for LCLS data.]{
            \scalebox{1}{
                \begin{tikzpicture}[node distance=1.1cm]
                    \node (data_in) [] {Input};
                    \node (fc1) [nn, below of=data_in] {Fully-connected + ReLU};
                    \node (fc2) [nn, below of=fc1] {Fully-connected + ReLU};
                    \node (fc3) [nn, below of=fc2, xshift=-3cm, yshift=-0.5cm, text width=4cm] {Fully-connected};
                    \node (fc4) [nn, below of=fc2, xshift=3cm, yshift=-0.5cm, text width=4cm] {Fully-connected + Tanh w/ temp=20};
                    \node (mu) [below of=fc3] {$\mu$};
                    \node (std) [below of=fc4] {$\sigma$};
            
                    \draw [arrow] (data_in) -- (fc1) node [pos=0.5, right] {453};
                    \draw [arrow] (fc1) -- (fc2) node [pos=0.5, right] {512};
                    \draw [arrow] (fc2) -- (fc3) node [pos=0.5, left] {512};
                    \draw [arrow] (fc2) -- (fc4) node [pos=0.5, right] {512};
                    \draw [arrow] (fc3) -- (mu) node [pos=0.5, right] {64};
                    \draw [arrow] (fc4) -- (std) node [pos=0.5, right] {64};
                \end{tikzpicture}
            }
            \label{fig:lcls_arch_enc}
        }
        \subfigure[Decoder architecture for LCLS data.]{
            \scalebox{1}{
                \begin{tikzpicture}[node distance=1.1cm]
                    \node (data_in) [] {Input};
                    \node (fc1) [nn, below of=data_in] {Fully-connected + ReLU};
                    \node (fc2) [nn, below of=fc1] {Fully-connected + ReLU};
                    \node (fc3) [nn, below of=fc2, xshift=-3cm, yshift=-0.5cm, text width=4cm] {Fully-connected};
                    \node (fc4) [nn, below of=fc2, xshift=3cm, yshift=-0.5cm, text width=4cm] {Fully-connected + Tanh w/ temp=20};
                    \node (mu) [below of=fc3] {$\mu$};
                    \node (std) [below of=fc4] {$\sigma$};
            
                    \draw [arrow] (data_in) -- (fc1) node [pos=0.5, right] {67};
                    \draw [arrow] (fc1) -- (fc2) node [pos=0.5, right] {512};
                    \draw [arrow] (fc2) -- (fc3) node [pos=0.5, left] {512};
                    \draw [arrow] (fc2) -- (fc4) node [pos=0.5, right] {512};
                    \draw [arrow] (fc3) -- (mu) node [pos=0.5, right] {450};
                    \draw [arrow] (fc4) -- (std) node [pos=0.5, right] {450};
                \end{tikzpicture}
            }
            \label{fig:lcls_arch_dec}
        }
        \caption{\label{fig:lcls_arch} Model architecture for ResVAE on the LCLS data.}
    \end{figure}
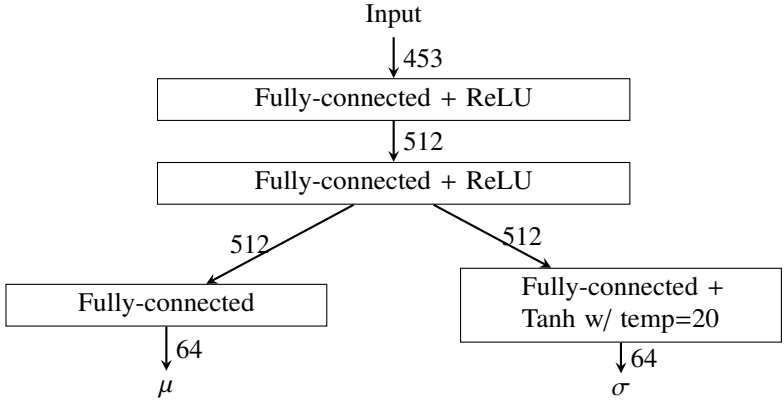
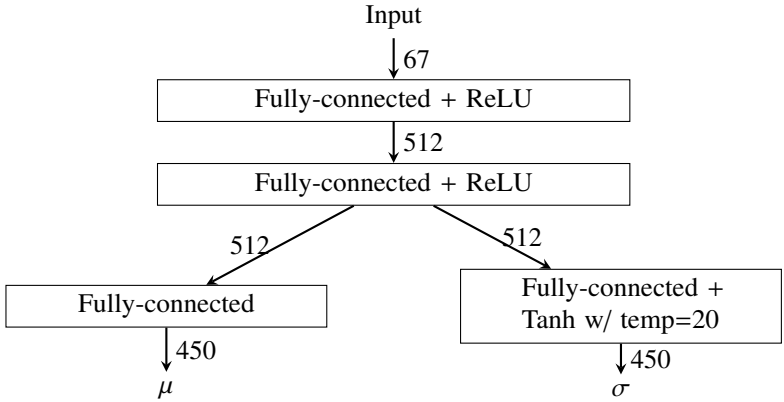

    \subsection{Anomaly figures}
    \cref{fig:beam_loss_signal} shows the TMIT signals, relative to the injector, along the beam for the period of anomalous behavior corresponding to~\cref{fig:beam_loss}. \cref{fig:x_anom_signal} shows the X signals along the beam for the period of anomalous behavior corresponding to~\cref{fig:x_anom}. \cref{fig:tmit_anom_signal} shows the TMIT signals, relative to the injector, along the beam for the period of anomalous behavior corresponding to~\cref{fig:tmit_anom}.

    \begin{figure}[htbp]
        \centering
        \includegraphics[width=\textwidth]{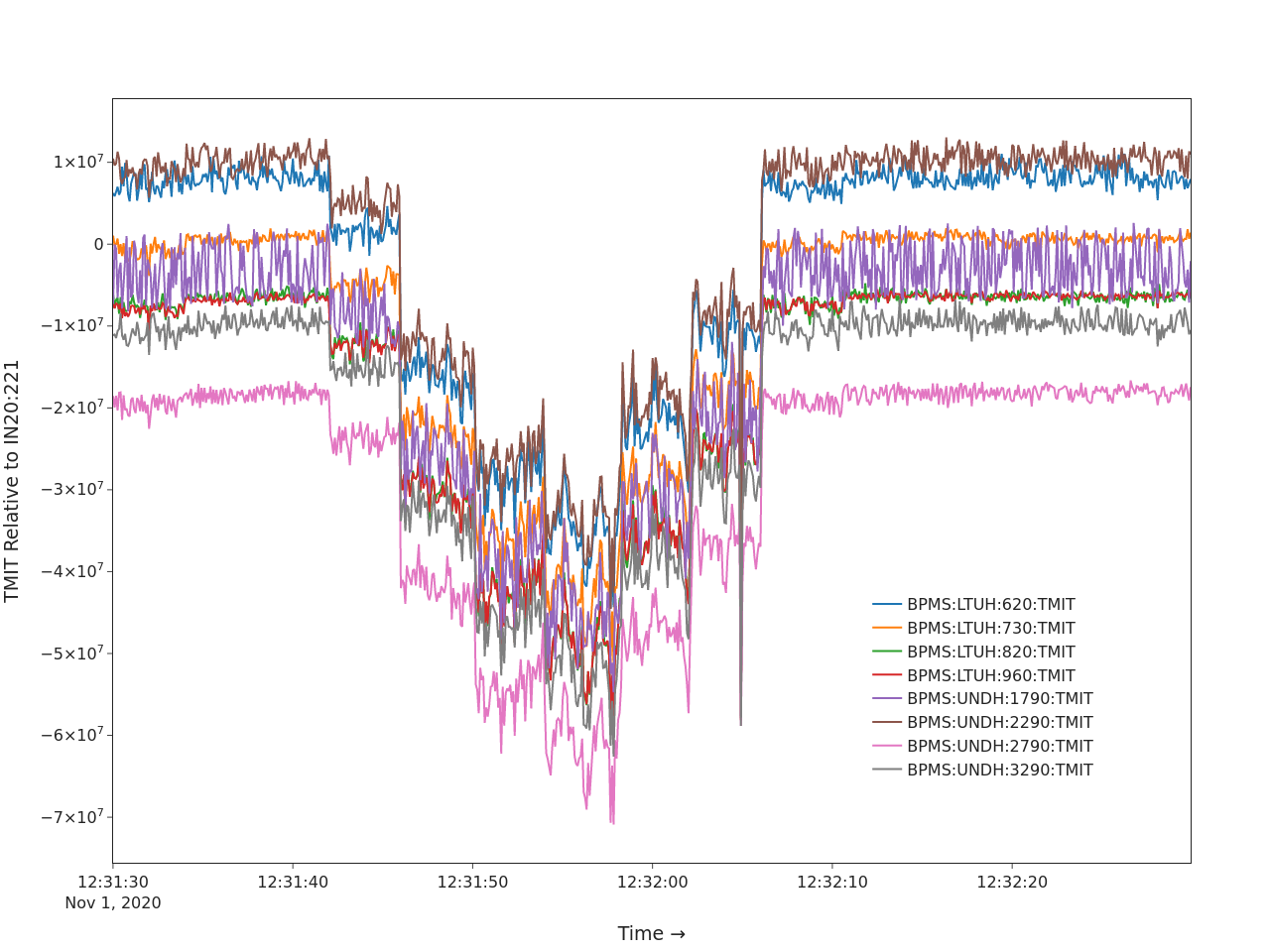}
        \caption{Plot of the TMIT (transmitted intensity) along the beam relative to the injector (IN20:221), showing a period of beam loss. The dip corresponds to beam loss. Corresponds to~\cref{fig:beam_loss}.}\label{fig:beam_loss_signal}
    \end{figure}

    \begin{figure}[htbp]
        \centering
        \includegraphics[width=\textwidth]{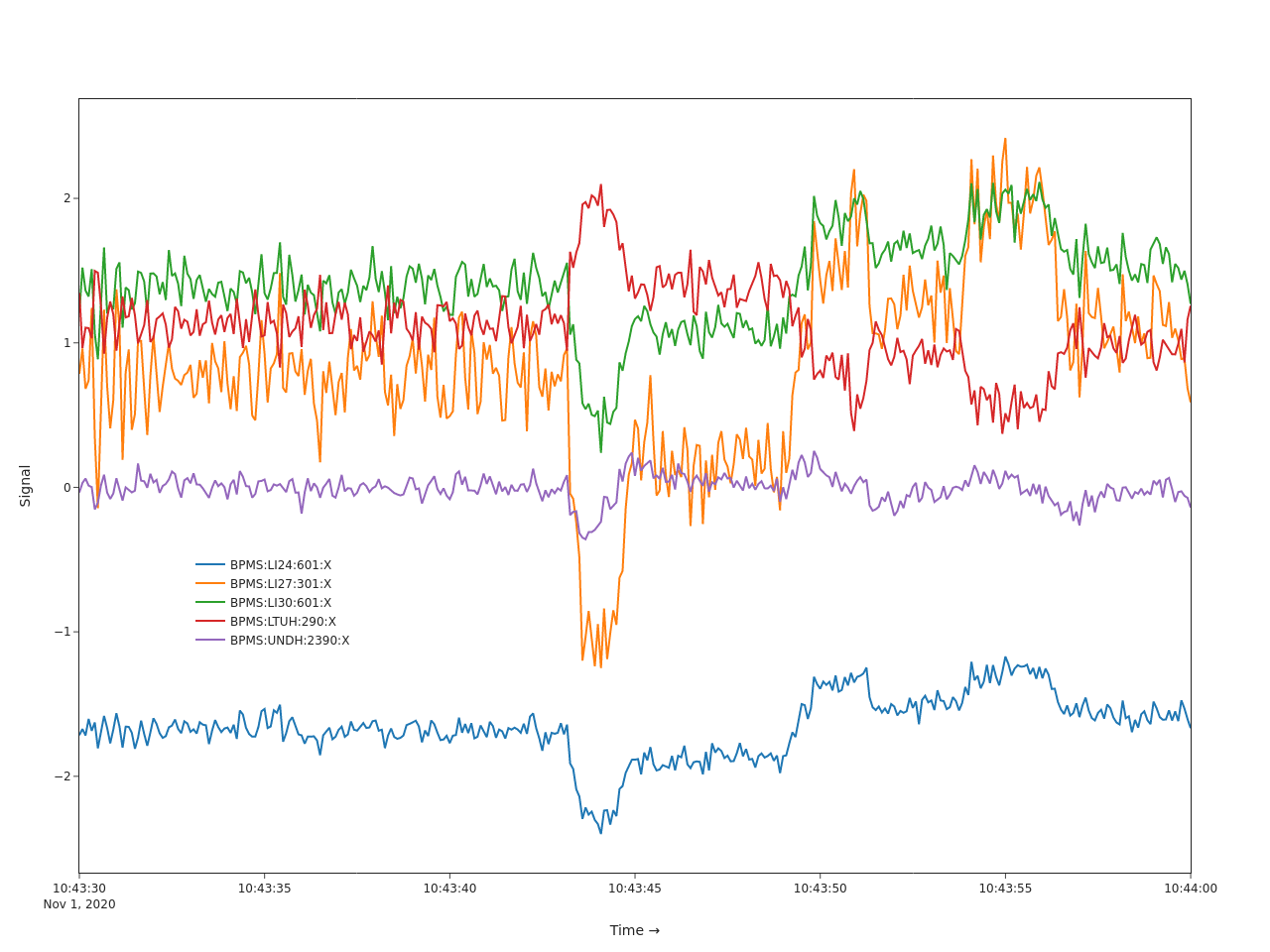}
        \caption{Plot of a subset of the (normalized) X signals along the beam, showing a period of anomalous behavior. Corresponds to~\cref{fig:x_anom}.}\label{fig:x_anom_signal}
    \end{figure}

    \begin{figure}[htbp]
        \centering
        \includegraphics[width=0.7\textwidth,trim={0 0 0 2cm},clip]{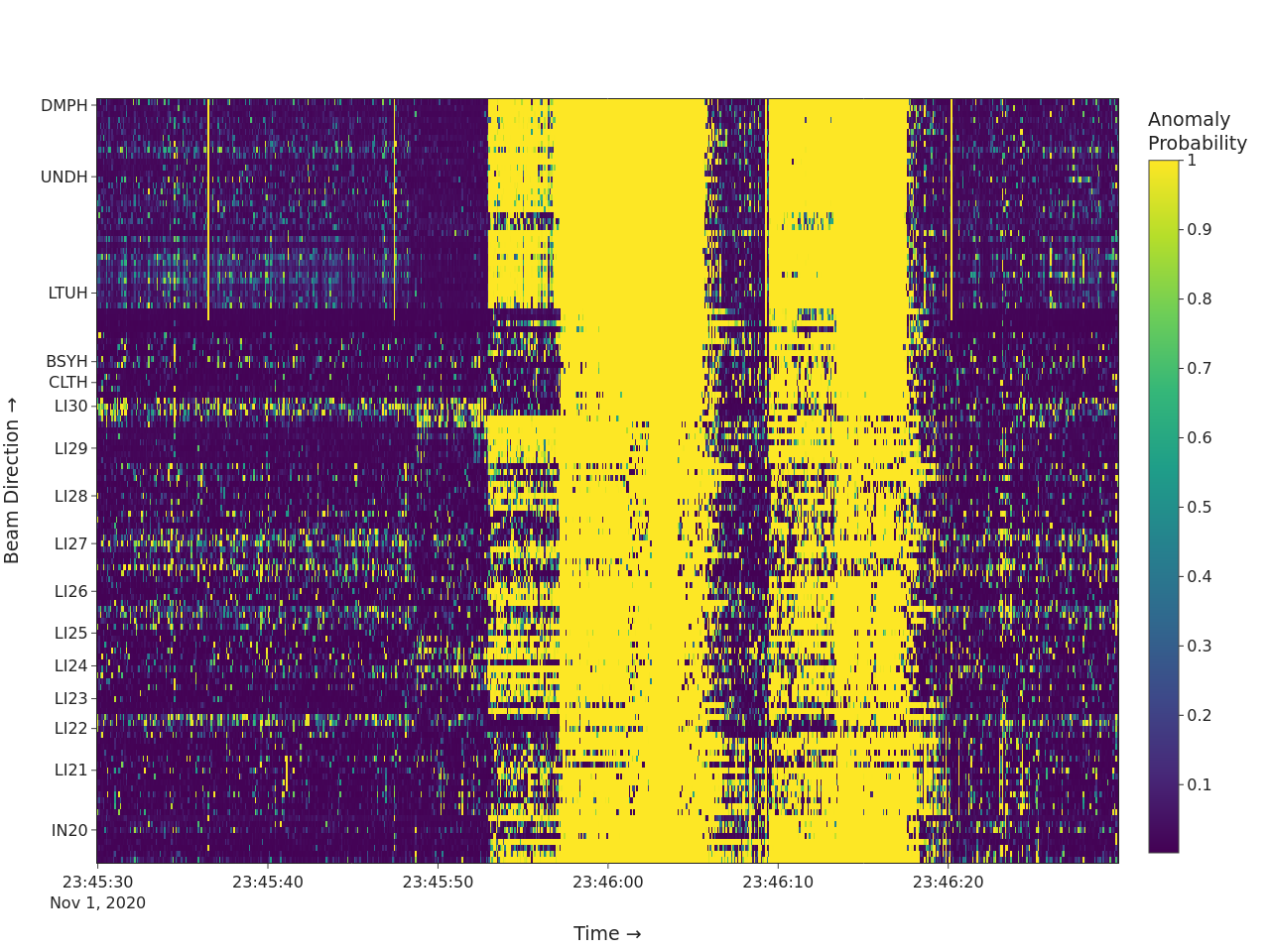}
        \caption{Example of anomaly detected by ResVAE and its feature-level anomaly attribution (where the feature-level anomaly probability is $ 1 - \pi_d(x) $). We only show the TMIT signals. We plot an illustrative subset of these signals in~\cref{fig:tmit_anom_signal}.}\label{fig:tmit_anom}
    \end{figure}

    \begin{figure}[htbp]
        \centering
        \includegraphics[width=\textwidth]{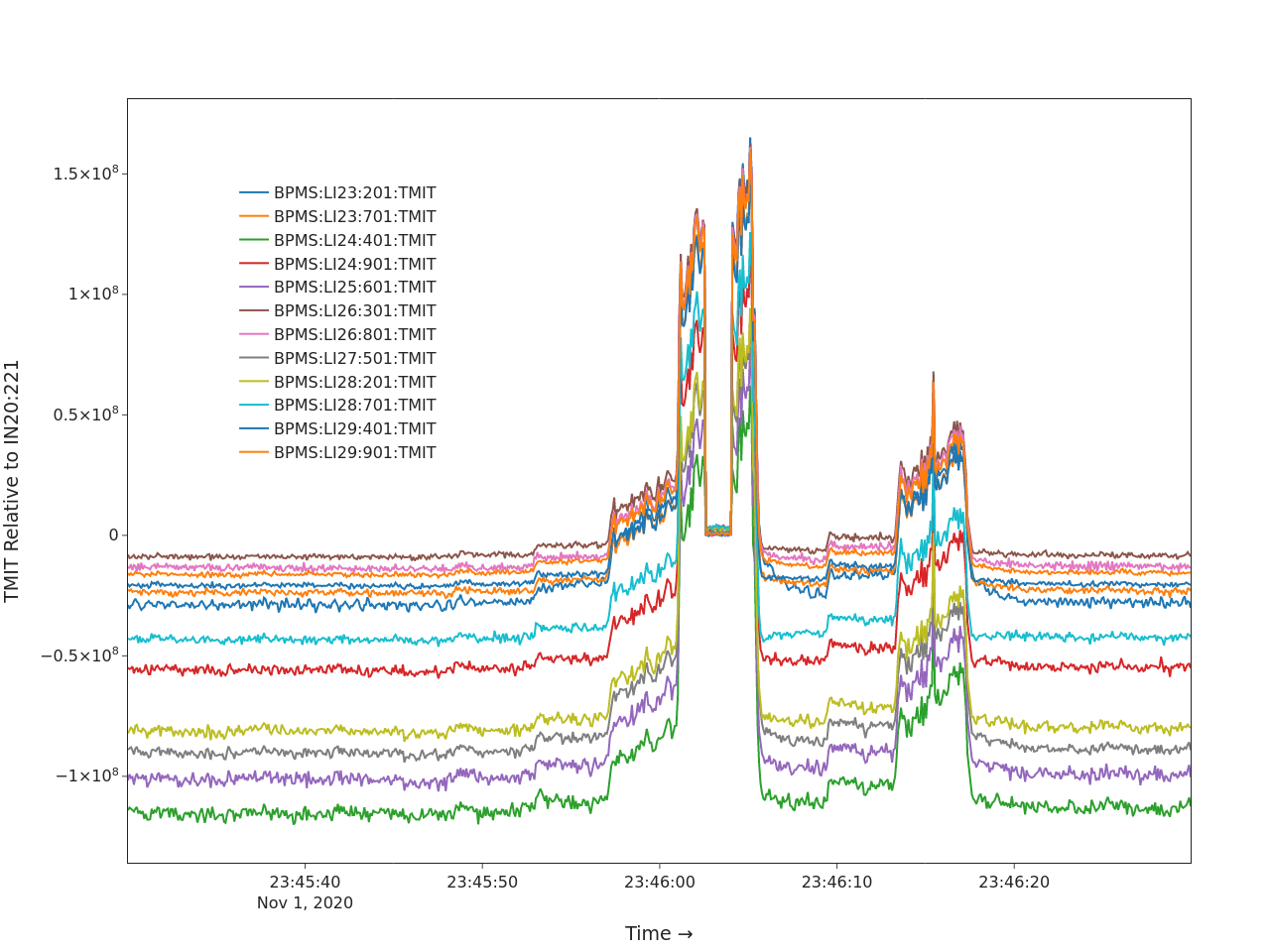}
        \caption{Plot of the TMIT (transmitted intensity) along the beam relative to the injector (IN20:221), showing a period of anomalous beam charge behavior. The flat-line section near the middle of the plot is total beam loss. Corresponds to~\cref{fig:tmit_anom}.}\label{fig:tmit_anom_signal}
    \end{figure}
\end{toappendix}

\section{Related Work}\label{sec:related_work}

There are numerous prior works studying the problem of unsupervised anomaly detection. Classical methods~\cite{Reynolds09_GMM,Parzen62_KDE,Scholkopf01_OCSVM,Liu08}, typically rely on density or distance and therefore struggle with high-dimensional inputs, especially when anomalies are present in the training set.
The main alternative methods rely on compression techniques, such as autoencoders~\cite{Chen17,Lu17_OD_AE}, VAEs~\cite{Wang17_VAE}, and generative adversarial nets (GANs)~\cite{Liu19_GAAL,Su19_OmniAnomaly}, or combine deep compression with classical methods~\cite{Zong18_DAGMM,Ruff18_DSVDD}. Nonetheless, these methods almost always assume a normal training set to avoid learning to compress anomalous examples\showto{thesis}{ and thereby distinguish normal and anomalous examples}.
 
Several methods attempt to add robustness to outliers to existing methods, including \showto{thesis}{robust KDE~\cite{Kim08_RKDE}, }robust PCA~\cite{Candes11_RPCA, Zhou17_DeepRPCA}, $\beta$-robust VAE~\cite{Akrami22_RobustVAE_Div}, and RVAE~\cite{Eduardo20_RVAE}. However, all of these approaches have various drawbacks. Although robust PCA (and its deep counterpart) can effectively identify both sample and feature anomalies in contaminated training data, its main drawbacks are the relative difficulty of interpreting the hyperparameter controlling the outlier fraction and the need to store the outlier matrix $ S $ of size equal to the dataset.
$\beta$-robust VAE~\cite{Akrami22_RobustVAE_Div} replaces the reconstruction likelihood term of the traditional ELBO with a more robust $\beta$-divergence term; however, this does not address robustness concerns in the latent space nor does this enable the model to identify feature outliers. Our approach is most similar to and extends the RVAE~\cite{Eduardo20_RVAE} work, following its general structure and derivation but making several key changes (detailed in~\cref{sec:rvae_compare}) to add robustness to the latent space and ensure expressive weights.

\begin{toappendix}
    \section{Comparison to RVAE}\label{sec:rvae_compare}
    First, we introduced additional mixing in the latent space with the mixing variable $ v $ (we recover the RVAE model if $ \rho = 1 $). Without this additional mixing, the encoder $ q_\phi(z|x) $ is forced to match both inlier and outlier examples to the same latent prior $ p(z) $, but as we argued in~\cref{sec:vae}, the prior-matching penalty is not robust to outliers. Second, we changed the definition of the outlier distributions, preserving the motivation of providing an outlier reference distribution but using the formalized $\delta$-diffuse definition for an exponential family distribution. Lastly, we replaced the ELBO objective with a constrained-optimization objective in order to introduce the regularization coefficients $ \beta_1, \beta_{2,d}, \beta_3 $ and interpret the functions $ \pi_d(x), \gamma(x) $ as logistic regression subproblems, which we solve with a pseudo-labeling approach. Without this reformulation, the functions $ \pi_d(x), \gamma(x) $ are not expressive (i.e., do not map inliers near $ 1 $ and outliers near $ 0 $).
\end{toappendix}

\begin{toappendix}
    \begin{Section}{thesis}{Synthetic Dataset Experiment}\label{sec:synthetic_dataset}

    \subsection{Dataset setup}
    In order to illustrate our method under controlled conditions, we create a synthetic outlier dataset. We first generate the inliers from $ \mathcal{N}\left(0, \Sigma \right) $ where $ \Sigma = U U^T $ and $ U \in \mathbb{R}^{64 \times 16} $ is the first $ 16$ columns of a random $64 \times 64$ orthogonal matrix. We then introduce outliers with a controllable difficulty $ \tau $ such that the dataset is only a fraction $ f $ inlier. Following the synthetic outlier construction in~\cite{Han22_ADBench}, we introduce four types of outliers:
    \begin{enumerate}
        \item \textit{Local}: Follows dependency structure but falls outside of inlier neighborhood. Drawn from $ \mathcal{N}\left(0, \alpha \Sigma\right) $ where $ \alpha $ depends on $ \tau $.
        \item \textit{Global}: Draws most features from the truncated standard Gaussian over $ [-0.7, 0.7] $ but some from $ [-10, -2] \cup [2, 10] $. The replacement rate depends on $ \tau $.
        \item \textit{Dependency}: Breaks dependency structure by generating some features according to the individual marginal distributions $\mathcal{N}\left(0, \Sigma_{d,d} \right)$ and the rest from the remaining conditional distribution. The replacement rate depends on $ \tau $.
        \item \textit{Clustered}: Samples from another cluster $ \mathcal{N}\left(\mu_a, \Sigma_a\right)$ where $ \mu_a $ is the center of the new cluster and $ \Sigma_a = U_a U^T_a $ for a different orthogonal matrix $ U_a \in \mathbb{R}^{64 \times 16} $. $ \mu_a $ depends on $ \tau $.
    \end{enumerate}
    We consider five difficulty levels $ \tau \in [0.01, 0.04, 0.07, 0.1] $, five outlier configurations (the four types individually plus an even mix of the four types), and three training inlier fractions $ f \in [0.8, 0.9, 1] $. 

    \subsection{Results}
    Across these variants of the dataset, we compare our method to the vanilla VAE~\cite{Kingma2014_VAE} and RVAE~\cite{Eduardo20_RVAE} using area under the precision-recall curve (AUPRC)~\cite{Davis06_AUPRC} as our metric. \cref{tab:blob_improvements} shows the AUPRC improvement compared to a vanilla VAE for the different outlier configurations, where we average across the different difficulty levels and training inlier fractions and present the best results for RVAE and ResVAE across a hyperparameter sweep. RVAE does not on average improve over a vanilla VAE, whereas our method ResVAE offers significant AUPRC improvements across all outlier configurations. \cref{tab:blob_improvements} also shows the results of an ablation study on our contributions. Starting from RVAE, we first swap in our diffuse outlier distributions (ResVAE w/o Lat. \& Logist.), then add the latent space mixing (ResVAE w/o Logist.), and finally add the logistic learning component to complete our full method. The ablation helps reveal the source of the AUPRC improvement; in particular, the improvement depends on the type of outlier. The diffuse outlier distributions yield improvements for \textit{global}, \textit{dependency}, and \textit{clustered} outlier types; the latent mixing yields additional improvement for \textit{clustered} anomalies; and the logistic learning yields improvements for each outlier configurations.
    
    \begin{table}[htbp]
        \centering
        \caption{Improvement in AUPRC over vanilla VAE, averaged across difficulty levels and training inlier fractions. \textit{All} denotes an equal mix of the four outlier types. In addition to our full ResVAE method, two intermediate variants are shown for ablation.}\label{tab:blob_improvements}
        \begin{tabular}{@{}llllll@{}}
        \toprule
                                     & All & Local & Global & Dependency & Clustered \\ \midrule
        VAE~\cite{Kingma2014_VAE} & 0 & 0 & 0 & 0 & 0 \\
        RVAE~\cite{Eduardo20_RVAE}   & 0   & +0.01  & +0.01   & 0          & -0.01   \\
        ResVAE w/o Lat. \& Logist.   & 0   & 0  & +0.08   & +0.10          & +0.10   \\
        ResVAE w/o Logist.           & +0.01   & 0  & +0.08   & +0.11          & +0.22   \\
        ResVAE                       & +\textbf{0.1} & +\textbf{0.05}  & +\textbf{0.14}   & +\textbf{0.18}       & +\textbf{0.28}    \\ \bottomrule
        \end{tabular}
    \end{table}
    
    We also consider the impact of different settings of $ \rho $ and $ \alpha_d $, which are ResVAE's primary hyperparameters and control the prior assumption on degree of dataset contamination, by sweeping across $ \rho, \alpha_d \in [0.8, 0.9, 1] $. We find that the best overall AUPRC, across all difficulty levels and training inlier fractions, is achieved by $ \rho = \alpha_d = 0.8 $; this configuration also gives the best AUPRC on \textit{local} and \textit{dependency} outlier configurations. The best AUPRC for \textit{global} outliers was achieved by $ \rho = \alpha_d = 0.9 $. Lastly, $ \rho = 0.8, \alpha_d = 1 $ gave the best AUPRC for \textit{clustered} outliers; since clustered outliers can be compressed and reconstructed but only through a different latent representation, we would expect the latent space (instead of feature reconstruction) to provide the discrimination between inliers and outliers in this case.
    \end{Section}
\end{toappendix}

\section{Conclusion}\nosectionappendix

This study introduces the ResVAE anomaly detection method, an innovative extension of the VAE model~\cite{Kingma2014_VAE} designed to detect anomalies even when trained on contaminated data. We provide a theoretical explanation for the limitations of VAE when faced with contaminated training data and propose a solution using a mixture model that assigns weights to both samples and features during training, effectively mitigating the impact of outliers. Subsequently, we apply this method to anomaly detection at LCLS, successfully identifying anomalous accelerator status and illustrating various types of anomalies.

\bibliography{ResVAE}

\end{document}